\documentclass[preprint,prl,showpacs,nofootinbib]{revtex4}
\usepackage{graphicx}
\usepackage{amsmath}
\usepackage{subfigure}

\begin{document}

\title{Magnetosonic solitons in a dusty plasma slab}

\author{M.\ Marklund}
\email{mattias.marklund@physics.umu.se}
\affiliation{Department of Physics, Ume{\aa} University, SE--901 87
  Ume{\aa}, Sweden} 

\author{L.\ Stenflo} 
\affiliation{Department of Physics, Ume{\aa} University, SE--901 87
  Ume{\aa}, Sweden} 

\author{P.K.\ Shukla}
\affiliation{Department of Physics, Ume{\aa} University, SE--901 87
  Ume{\aa}, Sweden}

\date{Received July 22, 2007}

\begin{abstract}
The existence of magnetosonic solitons in dusty plasmas is investigated.  
The nonlinear magnetohydrodynamic equations for a warm dusty 
magnetoplasma are thus derived. A solution of the nonlinear
equations is presented. It is shown that, due to the presence of dust, 
static structures are allowed. This is in sharp contrast to the formation of the so 
called shocklets in usual magnetoplasmas. A comparatively small number of dust particles can thus 
drastically alter the behavior of the nonlinear structures in magnetized plasmas.
\end{abstract}
\maketitle

In previous studies (e.g. Refs. [1, 2, 3]) of nonlinear magnetohydrodynamic waves, it has been shown that 
magnetosonic waves can appear both as solitary waves and as shock waves in different kinds of plasmas [4--6]. 
The formation of magnetosonic shocklets in space plasmas has been discussed in recent papers [7]. 
For perturbations traveling across the external magnetic field direction, it was furthermore shown 
that exact analytical solutions of the magnetohydrodynamic (MHD) equations can be found for a cold plasma [5]. 
The nonlinear solutions have also been obtained for a warm magnetoplasma [6], as well as for investigating matter 
waves in dilute gases [8]. Later more general [9], or alternative [10], solutions were also presented.

In the present paper we are going to deduce a nonlinear model for fast magnetosonic solitons
(FMS)  
in a plasma where a layer of \emph{charged dust} particles [11] is also present. It will then turn out that 
the presence of dust particles can significantly change the results of previous papers [2, 5] where 
electron-ion plasmas without dust were considered. Dusty plasmas can appear in many contexts, for example 
in interplanetary space, in cometary tails and comae, in planetary rings, in the Earth's atmosphere, 
as well as in dc and rf discharges, in plasma processing reactors, in solid-fuel combustion products, 
and in fusion plasma devices [12], and dusty layers [13] exist in the Earth's mesosphere \cite{havnes}.

The dynamics of the nonlinear FMS waves in a dusty magnetoplasma is governed by 
a set of equations composed of inertialess electron momentum equation
\begin{equation}
0 = -n_e e {\bf E} - \nabla p_e -n_e e\frac{{\bf v}_e}{c}\times {\bf B},
\label{eq1}
\end{equation}
the ion continuity equation
\begin{equation}
\frac{\partial n_i}{\partial t} + \nabla \cdot (n_i {\bf v}_i) =0,
\label{e2}
\end{equation}
and the ion momentum equation
\begin{equation}
m_i n_i \left( \frac{\partial}{\partial t}
+ {\bf v}_i \cdot \nabla\right) {\bf v}_i = n_i e {\bf E}-\nabla p_i 
+ n_i e \frac{{\bf v}_i}{c}\times {\bf B},
\label{e3}
\end{equation}
where $n_e (n_i) $ is the electron (ion) number density, $e$ is the magnitude of the
electron charge, ${\bf E}$ is the wave electric field, ${\bf B}$
is the sum of the ambient and wave magnetic fields, $p_e (p_i)  =n_e T_e (n_i T_i)$ 
is the electron pressure, $T_e (T_i)$ is the electron (ion) temperature, ${\bf v}_e ({\bf v}_i)$
is the electron (ion) fluid velocity, $m_i$ is the ion mass, and $c$ is the speed of light in  vacuum.
Following standard theory, we consider the plasma as isothermal, i.e. $T_e$  and $T_i$ are constants.
 Equations (1)-(3) are supplemented by means of Amp\`ere's law 
\begin{equation}
\nabla \times {\bf B} =\frac{4\pi e}{c}(n_i {\bf v}_i -n_e {\bf v}_e),
\label{e4}
\end{equation}
and Faraday's law
\begin{equation}
\frac{\partial {\bf B}}{\partial t} =-c \nabla \times {\bf E},
\label{e5}
\end{equation}
together with the quasi-neutrality condition $n_e =n_i - Z_dn_d$. Here
$n_d(x)$ is the prescribed number density of the dust, while $Z_d$ is the
number of electrons on each dust grain. 
As the dust particles are assumes to be very heavy, we can consider them as immobile. 
Collisional effects are also neglected.  
The quasi-neutrally condition holds for a dense plasma in which the ion plasma
frequency is much larger than the ion gyrofrequency. Equation (4) is valid
for FMS waves whose phase speed is much smaller than the speed of light.

Eliminating ${\bf E}$ from (1) and (3) we obtain \cite{shu03}
\begin{equation}
\left(\frac{\partial}{\partial t} + {\bf v}_i \cdot \nabla \right){\bf v}_i 
 = -\frac{1}{m_i}\left( \frac{T_e}{n_i - Z_dn_d} + \frac{T_i}{n_i}
 \right)\nabla n_i %+  \frac{Z_dT_e}{m_i(n_i - Z_dn_d)}\nabla n_d  
 + \frac{(\nabla \times {\bf B}) \times {\bf B} }{4\pi m_i(n_i - Z_dn_d)} 
%\nonumber \\ && 
 - \frac{eZ_dn_d}{cm_i(n_i - Z_dn_d)}{\bf v}_i\times{\bf B} .
 \label{e6}
 \end{equation}
On the other hand, from (1), (4) and (5) we have
\begin{equation}
\frac{\partial {\bf B}}{\partial t} = 
\nabla \times \left[ \frac{n_i}{n_i - Z_d n_d}{\bf v}_i \times {\bf
    B} -\frac{c}{4\pi e (n_i - Z_dn_d)}(\nabla \times {\bf B})\times {\bf
    B}\right]. 
\label{e7}
\end{equation}

We are interested in studying the nonlinear properties of one-dimensional
FMS waves across the external magnetic field direction $\hat {\bf z} B_0$,
where $\hat {\bf z}$ is the unit vector along the $z$ axis and $B_0$ is the
strength of the ambient magnetic field. Thus, we have $\nabla =\hat {\bf x}
\partial/\partial x$, ${\bf v}_i =u(x) \hat {\bf x} + w(x) \hat{\bf
  y}$ and ${\bf B} =B(x)\hat {\bf z}$, where $\hat {\bf x}$ ($\hat{\bf
  y}$) is the unit vector along the $x$ ($y$) axis  
in Cartesian coordinates. Thus, with these prerequisites, the MHD
equations become
\begin{subequations}
\begin{equation}
  \frac{\partial n}{\partial t} + \frac{\partial}{\partial x}(nu) = 
  0, 
\end{equation}
\begin{equation}
  \left(\frac{\partial}{\partial t} + u\frac{\partial}{\partial
  x}\right)u = -\frac{1}{m}\left( \frac{T_e}{n - Z_dn_d} + \frac{T}{n}
 \right)\frac{\partial n}{\partial x} -
\frac{1}{4\pi m (n - Z_dn_d)}B\frac{\partial
  B}{\partial x} 
%\nonumber \\ && 
 - \frac{eZ_dn_d}{m (n - Z_dn_d)}\frac{w}{c}B, 
\end{equation}
\begin{equation}
  \left(\frac{\partial}{\partial t} + u\frac{\partial}{\partial
  x}\right)w =  \frac{eZ_dn_d}{m (n - Z_dn_d)}\frac{u}{c}B , 
\end{equation}
and 
\begin{equation}
\label{eq:mag}
  \frac{\partial B}{\partial t} = - \frac{\partial}{\partial x}\left(
  \frac{n}{n - Z_d n_d}uB \right) ,
\end{equation}
\label{eq:system}
\end{subequations}
where we have dropped the index $i$. We note that the magnetic field is frozen-in 
according to 
\begin{equation}
  B/B_0 = n_e/n_0 , 
\end{equation}
for some constant $n_0$, 
and that $w \rightarrow 0$ as $n_d \rightarrow 0$. 
However, these conditions are modified by the inclusion of dust species, leaving
$w \neq 0$ in the generic case. Thus, the inclusion of dust breaks the frozen-in-field line
symmetry between the electron and ion species, similar to the effect of including 
inertial terms in a non-dusty plasma. As will be seen below, this has profound effects on
the nonlinear dynamics of magnetized dusty plasmas.  

Next, we look for stationary solutions, i.e. $\partial_t = 0$.
With this, the above equations can be integrated. The continuity equation yields the 
velocity component $u$ according to 
\begin{equation}
  u(x) = \frac{u_0n_0}{n(x)},
\end{equation}
where $u_0$ is some constant of integration determined by our boundary conditions. 
For the velocity component $w$ we have the equation
\begin{equation}\label{eq:w}
  \frac{dw}{dx} = \frac{e}{mc}\frac{Z_dn_dB}{(n - Z_dn_d)} ,
\end{equation}
while from the momentum equation in the $x$-direction we obtain
\begin{eqnarray}
  \frac{d}{dx}\left[
    \frac{mu^2}{2} + T_e\ln(n - Z_dn_d) + T\ln n
  \right] + \frac{1}{(n - Z_dn_d)}\frac{d}{dx}\left(
    \frac{B^2}{8\pi}
  \right)
  = -\frac{e}{c}\frac{Z_dn_dBw}{(n - Z_dn_d)} .
\label{eq:mom}
\end{eqnarray}
Thus, the magnetic field is given by the
quasi-neutrality condition and (9), the $x$-component of the velocity is given by 
(10), and it remains to solve (11) and (12) for $n$ and $w$. 
  We note that (\ref{eq:w}) can be written  in energy form according to
  $\partial_x(mw^2/2) = ({e}/{c})Z_dn_dBw/(n - Z_dn_d)$, and adding this to 
  Eq.\ (\ref{eq:mom}) gives $\partial_x[m(u^2 + w^2)/2 + T_e\ln n_e + T\ln n]
  + n_e^{-1}\partial_x(B^2/8\pi) = 0$.

Next we normalize our variables according to $n_d \rightarrow Z_dn_d/n_0$,
$n\rightarrow n/n_0$, $u \rightarrow u/u_0$, $w \rightarrow w/u_0$, $V_T \equiv (T/m)^{1/2}
\rightarrow V_T/u_0$, $V_A \equiv (B_0^2/4\pi mn_0)^{1/2} \rightarrow V_A/u_0$, 
$T_e \rightarrow T_e/T$, and $x \rightarrow x\omega_c/u_0$, where $\omega_c = eB_0/mc$
is the ion cyclotron frequency. 

Inserting (9) and (10) into (11) and integrating gives
\begin{equation}
  w(x) = \int_{-\infty}^x n_d(x')\,dx' ,
\end{equation}
while integration of (12) gives
\begin{equation}
  \frac{1}{2n^2} + \frac{w^2}{2} 
    + V_T^2\left[T_e\ln\left(n - n_d\right) + \ln n\right] 
    + V_A^2n = W_0 , 
\end{equation}
where $W_0$ is a constant. For a dust layer, we may use the density profile  
$n_d = (N_d/\sqrt{2\pi}\,a)\exp(-x^2/2a^2)$, where $N_d$ is the normalized number of dust particles 
and $a$ is the width of the dust slab. Such a dusty plasma slab admits solitary wave solutions.
In the figures, we show the numerical solutions to the normalized equations, using a gaussian dust density
distribution. In Fig.\ 1, we have used $a = 0.2$, $N_d = 1$, $V_T = 0.1$, $T_e = 10$, and $V_A = 10$, 
 in Fig.\ 2, we have used $a = 0.2$, $N_d = 1$, $V_T = 1$, $T_e = 100$, and $V_A = 10$, 
 and in Fig.\ 3, we have used $a = 0.2$, $N_d = 1$, $V_T = 10$, $T_e = 1$, and $V_A = 10$.
  In all cases, the velocity components stay bounded.

%%%%% FIG %%%%%
\begin{figure}
  \subfigure[]{\includegraphics[width=0.48\textwidth]{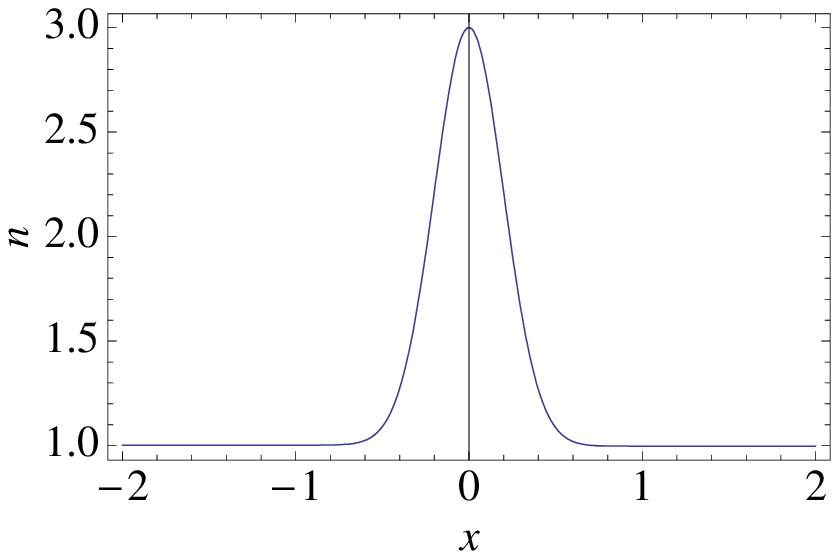}}
  \subfigure[]{\includegraphics[width=0.48\textwidth]{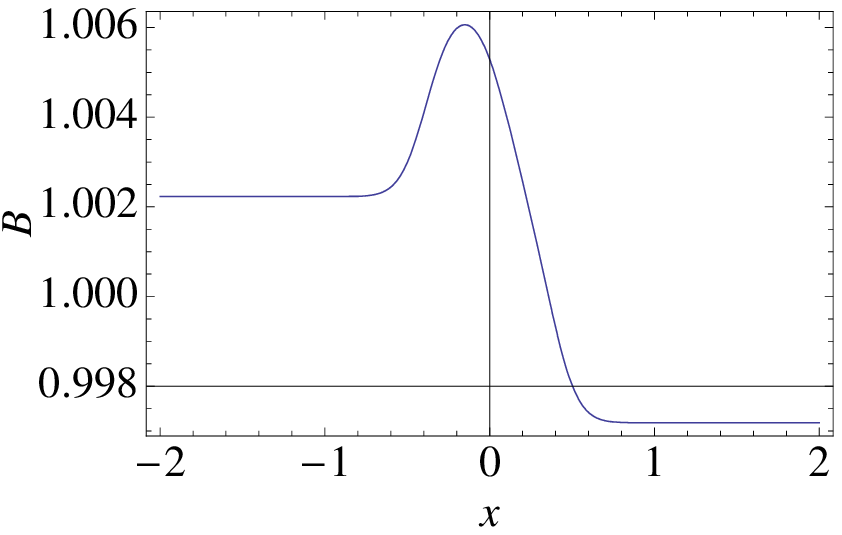}}
  \caption{The normalized ion density $n$ and magnetic field $B$
  plotted as a function of the normalized variable $x$ for
  $a = 0.2$, $N_d = 1$, $V_T = 0.1$, $T_e = 10$, and $V_A = 10$.}
\end{figure}
%%%%%%%%%%%%%

%%%%% FIG %%%%%
\begin{figure}
  \subfigure[]{\includegraphics[width=0.48\textwidth]{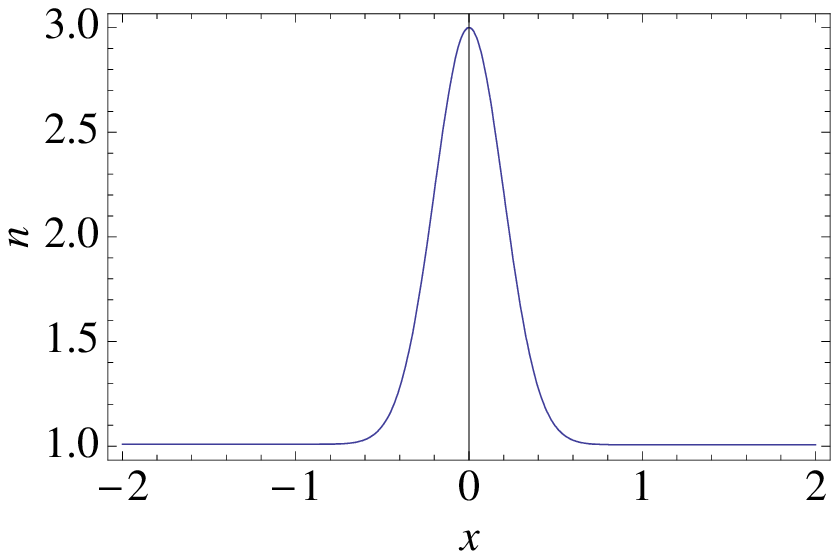}}
  \subfigure[]{\includegraphics[width=0.48\textwidth]{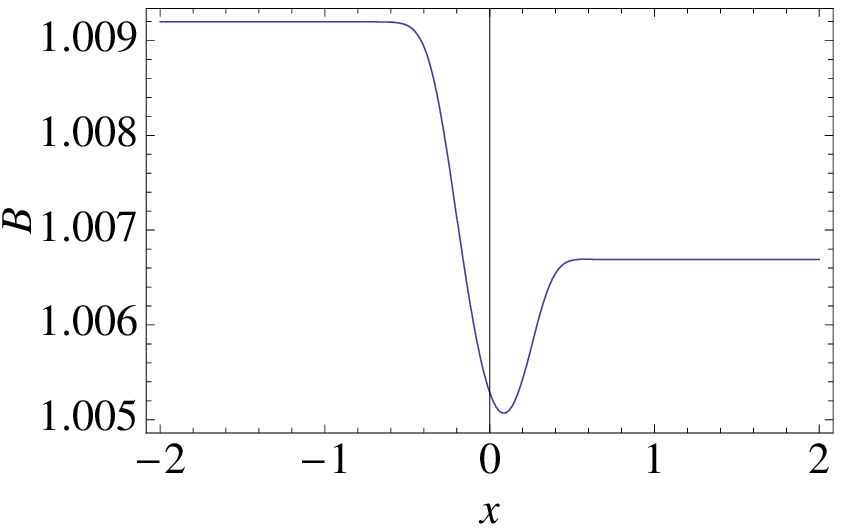}}
  \caption{The normalized ion density $n$ and magnetic field $B$
  plotted as a function of the normalized variable $x$ for
  $a = 0.2$, $N_d = 1$, $V_T = 1$, $T_e = 100$, and $V_A = 10$.}
\end{figure}
%%%%%%%%%%%%%

%%%%% FIG %%%%%
\begin{figure}
  \subfigure[]{\includegraphics[width=0.48\textwidth]{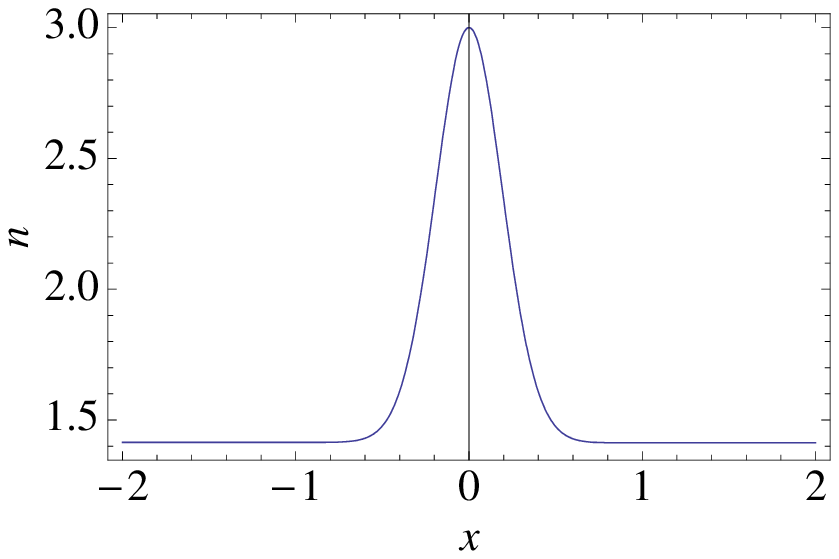}}
  \subfigure[]{\includegraphics[width=0.48\textwidth]{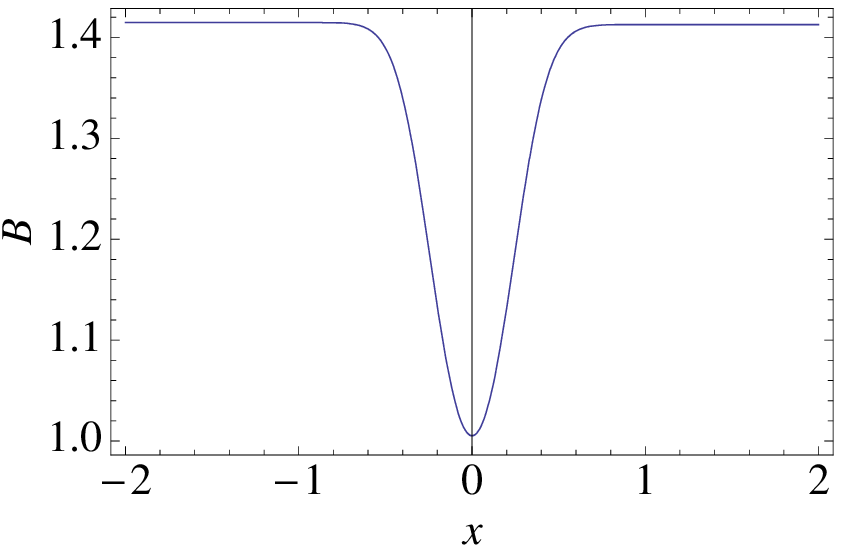}}
  \caption{The normalized ion density $n$ and magnetic field $B$
  plotted as a function of the normalized variable $x$ for
  $a = 0.2$, $N_d = 1$, $V_T = 10$, $T_e = 1$, and $V_A = 10$.}
\end{figure}
%%%%%%%%%%%%%

To summarize, we have presented an analytical theory for magnetosonic solitons 
in a magnetized electron-ion-dust plasma. It is found that the presence of a stationary 
charged dust layer provides the possibility 
of new classes of magnetosonic solitons. For typical dusty plasma
parameters, we have displayed the density and magnetic field profiles of compressional magnetosonic 
solitons. The present results may describe the salient features of localized 
magnetosonic solitons in forthcoming laboratory experiments in strong magnetic fields.

\end{document}